\documentclass[10pt,superscriptaddress,nofootinbib,twocolumn]{revtex4-2}

\usepackage{amsthm,amsmath,amssymb}
\usepackage{lineno}
\usepackage{graphicx}
\usepackage{subcaption}
\usepackage[usenames,dvipsnames]{color}
\usepackage[colorlinks=true,citecolor=blue,urlcolor=black]{hyperref}
\usepackage{dcolumn}
\usepackage{natbib}
\usepackage{booktabs}

\begin{document}
\title{Efficient source-independent quantum conference key agreement}

\author{Yu Bao}
\affiliation{National Laboratory of Solid State Microstructures and School of Physics, Collaborative Innovation Center of Advanced Microstructures, Nanjing University, Nanjing 210093, China}
\affiliation{Department of Physics and Beijing Key Laboratory of Opto-Electronic Functional Materials and Micro-Nano Devices, Key Laboratory of
Quantum State Construction and Manipulation (Ministry of Education), Renmin University of China, Beijing 100872, China}
\author{Yi-Ran Xiao}
\affiliation{National Laboratory of Solid State Microstructures and School of Physics, Collaborative Innovation Center of Advanced Microstructures, Nanjing University, Nanjing 210093, China}
\affiliation{Department of Physics and Beijing Key Laboratory of Opto-Electronic Functional Materials and Micro-Nano Devices, Key Laboratory of
Quantum State Construction and Manipulation (Ministry of Education), Renmin University of China, Beijing 100872, China}
\author{Yu-Chen Song}
\affiliation{Big Data Center, Ministry of Emergency Management, Beijing 100013, China}
\author{Xiao-Yu Cao}
\affiliation{National Laboratory of Solid State Microstructures and School of Physics, Collaborative Innovation Center of Advanced Microstructures, Nanjing University, Nanjing 210093, China}
\affiliation{Department of Physics and Beijing Key Laboratory of Opto-Electronic Functional Materials and Micro-Nano Devices, Key Laboratory of
Quantum State Construction and Manipulation (Ministry of Education), Renmin University of China, Beijing 100872, China}
\author{Yao Fu}
\affiliation{Beijing National Laboratory for Condensed Matter Physics and Institute of Physics, Chinese Academy of Sciences, Beijing 100190, China}
\author{Hua-Lei Yin}\email{hlyin@ruc.edu.cn}
\affiliation{Department of Physics and Beijing Key Laboratory of Opto-Electronic Functional Materials and Micro-Nano Devices, Key Laboratory of
Quantum State Construction and Manipulation (Ministry of Education), Renmin University of China, Beijing 100872, China}
\affiliation{Beijing Academy of Quantum Information Sciences, Beijing 100193, China}
\affiliation{Yunnan Key Laboratory for Quantum Information, Yunnan University, Kunming 650091, China}
\author{Zeng-Bing Chen}
\affiliation{National Laboratory of Solid State Microstructures and School of Physics, Collaborative Innovation Center of Advanced Microstructures, Nanjing University, Nanjing 210093, China}
\date{\today}

\begin{abstract}
Quantum conference key agreement (QCKA) enables the unconditional secure distribution of conference keys among multiple participants. Due to challenges in high-fidelity preparation and long-distance distribution of multi-photon entanglement, entanglement-based QCKA is facing severe limitations in both key rate and scalability. Here, we propose a source-independent QCKA scheme utilizing the post-matching method, feasible within the entangled photon pair distribution network. We introduce an equivalent distributing virtual multi-photon entanglement protocol for providing the unconditional security proof even in the case of coherent attacks. For the symmetry star-network, comparing with previous $n$-photon entanglement protocol, the conference key rate is improved from $O(\eta^{n})$ to $O(\eta^{2})$, where $\eta$ is the transmittance from the entanglement source to one participant. Simulation results show that the performance of our protocol has multiple orders of magnitude advantages in the intercity distance. We anticipate that our approach will demonstrate its potential in the implementation of quantum networks.
\end{abstract}
\maketitle

\section{INTRODUCTION}
The goal of the quantum network is to establish a quantum internet that will enhance the existing internet by allowing the collection, processing, storage and transmission of quantum data~\cite{simon2017towards,wehner2018quantum}. The development of quantum network is divided into six stages according to the functional division~\cite{wehner2018quantum}, some well-known cryptographic applications~\cite{cao2024experimental,schiansky2023demonstration,weng2023beating,PRXQuantum.4.020320,PhysRevX.10.011038} can be performed through quantum protocols at each stage. \textcolor{black}{Quantum cryptography, 
which plays a crucial role in quantum network, is moving towards practicality and provides confidentiality, integrity, authenticity and non-repudiation~\cite{yin2023experimental}. For instance, quantum privacy communication~\cite{lo2012measurement,PRXxie22,zeng2022mode,lucamarini2018overcoming,hu2016experimental,chen2021integrated} and quantum digital signature~\cite{PhysRevApplied.20.044011,dunjko2014quantum,yin2016practical,PhysRevX.11.011038} are basic quantum cryptography primitives, which can allow participants to implement data transmission with confidentiality and non-repudiation in large scalable situation, respectively. Quantum privacy communication includes quantum key distribution~\cite{zhou2023experimental,ghalaii2023satellite,lu2023experimental,gu2022experimental,PhysRevLett.131.110801,li2023twin}, quantum secret sharing~\cite{fu2015long,kogias2017unconditional,li2023breakingqss,senthoor2019communication,gu2021secure}, quantum secure direct communication~\cite{zhang2017quantum,qi202115,sheng2022one} and quantum conference key agreement (QCKA)~\cite{li2023breaking,pickston2023conference,liu2023experimental,bai2022post,hahn2020anonymous,li2021finite,carrara2023overcoming,proietti2021experimental,epping2017multi,murta2020quantum,zhao2020phase,cao2021coherent,grasselli2018finite}}. \textcolor{black}{Quantum} key distribution exploits two-party entanglement state, allowing two participants to share secure secret keys with the insecure quantum channels, thereby enabling peer-to-peer communication. QCKA is an extension of quantum key distribution to scenarios involving multiple participants. Faced with multi-party communication tasks, multi-photon entanglement-based QCKA allows for lower quantum and classical resource consumption~\cite{miguel2023optimized, wallnofer2019multipartite,li2023all,kuzmin2019scalable} compared to repeating quantum key distribution between every two participants and possesses greater development potential.

Multi-photon entanglement-based QCKA protocols require the generation of Greenberger-Horne-Zeilinger (GHZ) states~\cite{bose1998multiparticle,chen2005conference,grasselli2018finite,murta2020quantum,hahn2020anonymous} or ``twisted'' versions of them~\cite{mermin1990extreme,greenberger1989going,cai2018multipartite} corresponding to the number of participants. When employing such protocols, both the system complexity and error rate of entanglement source will significantly increase as the number of protocol participants rises. Several experimental studies~\cite{tittel2001experimental, proietti2021experimental, 
schmid2005experimental, gaertner2007experimental, erven2014experimental,pickston2023conference} have already analyzed the actual performance of the GHZ state distribution protocol, such as the experiment based on the ``N-BB84'' protocol~\cite{proietti2021experimental}, the N-partite version of the asymmetric BB84 quantum key distribution protocol. This work utilized state-of-the-art four-photon entanglement sources, encrypting and securely sharing an image among four participants over 50 km fiber. This kind of protocol is still far from being realized. Challenges such as preparing multi-photon entanglement state~\cite{jons2017bright, uppu2021quantum}, as well as issues like limited coincidence count rates, present significant obstacles.

A measurement-device-independent (MDI) QCKA protocol~\cite{fu2015long} was introduced to avoid the preparation and transmission of multi-photon entanglement states. It distributes post-selected GHZ entanglement, inspired by both the decoy state method and entanglement swapping operation. 
Subsequently, various protocols use weak coherent-state sources as a substitute for the entanglement source, which include schemes based on single-photon interference~\cite{cao2021high,bai2022post,cao2021coherent}. These types of protocol inherit features from related twin-field quantum key distribution~\cite{lucamarini2018overcoming}, which means strict request on the intensity and sending probability of signal states and decoy states. The optimal parameters change with both the number of participants and the distances between them. Moreover, these schemes rely on the independence of measurement devices and cannot be extended to scenarios involving more than three participants by only expanding system structures. Recently, an MDI-QCKA protocol based on spatial multiplexing and adaptive operation has been proposed~\cite{li2023breaking}, which has broken the limitation on key rate under multiple participants. However, in this work, confirming the arrival of transmitted photons requires quantum non-demolition measurements, and guiding these photons to the GHZ analyzer through optical switches necessitates high-speed active feedforward techniques. These technical requirements constrains its practical applications. The issues of poor practicality and low key rates in QCKA have not been completely addressed, particularly in scenarios involving multiple participants.

Quantum entanglement networks based on entangled photon pairs have achieved remarkable accomplishments through the utilization of spatial and wavelength multiplexing techniques ~\cite{wengerowsky2018entanglement,joshi2020trusted,liu202240}. Here, we propose a source-independent QCKA protocol based on the network of Bell states, which \textcolor{black}{features} straightforward hardware implementation and good scalability. With the aid of the post-matching method~\cite{lu2021efficient}, measurement outcomes among participants can be correlated through a series of classical operations (Note that post-matching method has been extended to construct asynchronous measurement-device-independent quantum key distribution~\cite{PRXxie22,zeng2022mode}). Thus, results equivalent to measurements of GHZ states can be obtained and the need for multi-photon entanglement sources is avoided. We employ an equivalent virtual protocol to prove that our scheme can resist coherent attacks. The simulation results and a composable finite-key analysis \textcolor{black}{are} also provided. The results indicate that our protocol achieves a transmission distance of over 320 kilometers under the condition of 6 participants. In the case of 3 participants, our protocol achieves a key rate three orders of magnitude higher than N-BB84~\cite{grasselli2018finite} at a transmission distance of 200 kilometers. As participant numbers increase, this key rate advantage becomes more evident. Furthermore, our protocol allows for freely adjusting the number of participants without altering the system structure of the entanglement provider and other protocol participants. The \textcolor{black}{system of our protocol} can be integrated into the complex and dynamic quantum networks. We believe that our scheme provides a promising method for providing keys to multiple participants with one entangled photon pair source.

\section{Protocol description}\label{sec2}

\begin{figure}[t!]
  \centering
  \includegraphics[width=1.0\columnwidth]{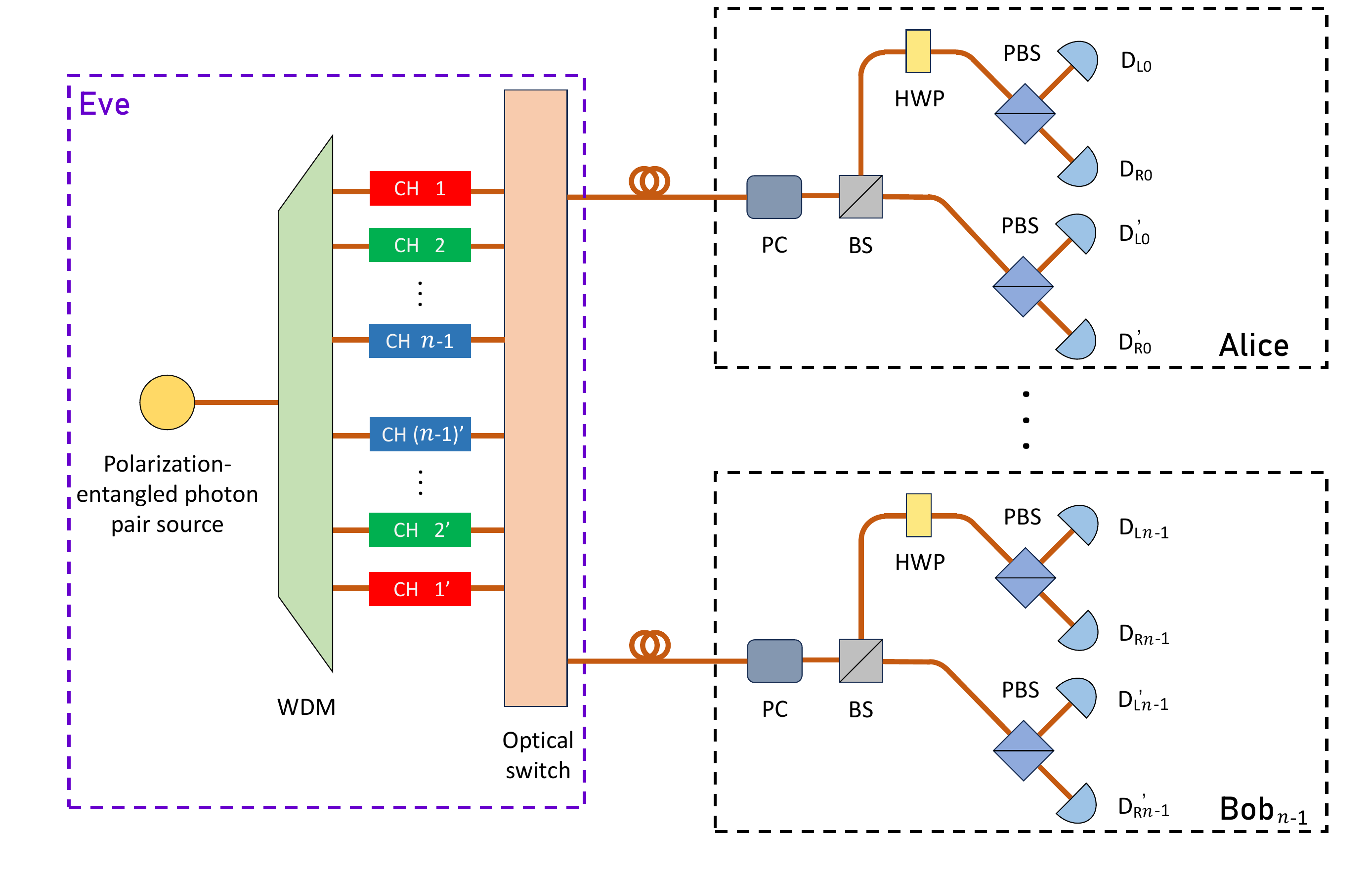}
  \caption{\textbf{The setup of QCKA protocol.}\label{fig1} The untrusted entanglement provider at the central node, named Eve, employs a polarization-entangled photon pair source to generate entangled photon pairs. These pairs can be distributed to $\rm Alice$ and $\rm Bob_i$ through the wavelength division multiplexer (WDM) and optical switch. $\rm Alice$ and $\rm Bob_i$ utilize a combination of polarization controller (PC), beam splitter (BS), half-wave plate (HWP), polarization beam splitter (PBS), and superconducting nanowire single-photon detector for their detection apparatus.}
\end{figure}

Our protocol is designed to be implemented based on entangled photon pairs distribution network. Any network user can be a protocol participant. The setup of our scheme, depicted in Fig.~\ref{fig1}, illustrates a centrally symmetric and source-independent design. To cover all wavelength channels effectively, we select a wavelength division multiplexer (WDM) based on the frequency range of the entanglement source. The generated Bell states, represented as $\left| \Phi^+ \right\rangle=\frac{1}{\sqrt{2}}(\left| 00 \right\rangle+\left| 11 \right\rangle)$ $(\left| \Phi^+ \right\rangle=\frac{1}{\sqrt{2}}(\left| ++ \right\rangle+\left| -- \right\rangle))$, can be allocated to different participants through WDM and optical switch~\cite{aktas2016entanglement, chang2016experimental, lim2008broadband}. $\left| 0 \right\rangle$, $\left| 1 \right\rangle$ are a pair of eigenstates of the Z basis, and $\left| + \right\rangle$, $\left| - \right\rangle$ are a pair of eigenstates of the X basis, where $\left| + \right\rangle=\frac{1}{\sqrt{2}}(\left| 0 \right\rangle+\left| 1 \right\rangle)$, $\left| - \right\rangle=\frac{1}{\sqrt{2}}(\left| 0 \right\rangle-\left| 1 \right\rangle)$. The measurement results for $\left| 0 \right\rangle$ and $\left| + \right\rangle$ are noted as 0, and for $\left| 1 \right\rangle$ and $\left| - \right\rangle$,  they are noted as 1. All participants have a set of devices to perform projection measurements on the received qubits.

While there exist $n$ participants named as $\rm Alice$, $\rm Bob_1$, $\rm Bob_2, \dots, \rm Bob_{\mathit{n}-1}$, the entire process of our protocol is presented as follows:

{\it{\bf{1.}}} $\rm Eve$ configures the optical switch according to the number and positions of the participants. Channel $\rm 1$, $\rm 2$, $\cdots$, $n-1$ are connected to $\rm Alice$ and Channel $\rm 1'$, $\rm 2'$, $\cdots$, $(n-1)'$ are connected to the corresponding \textcolor{black}{ $\rm Bob_\mathit{i} (\mathit{i}\in \{1, 2, \cdots, \mathit{n}-1\})$}. $\rm Eve$ prepares entangled photon pairs and distributes them to $\rm Alice$ and $\rm Bob_\mathit{i}$ via WDM and optical switch. The entangled photon pairs respectively pass through corresponding channels, $i$ and $i'$.

{\it{\bf{2.}}} $\rm Alice$ and $\rm Bob_\mathit{i}$ each select a basis and perform projection measurements on qubits, recording the outcomes and chosen basis. The probability of selecting the Z basis is represented as $p_a$ $(p_b)$ and the probability of selecting the X basis is $1-p_a$ $(1-p_b)$. All participants announce the bases they selected in the past period of time through classical channel in the agreed time slot. The events in which one entanglement state is both successfully measured by $\rm Alice$ and corresponding $\rm Bob_\mathit{i}$ under the same basis are denoted as valid events.

{\it{\bf{3.}}} The participants apply the post-matching method to handle valid events under the Z and X-bases. \textcolor{black}{$\rm Alice$ and $\rm Bob_\mathit{i}$ arrange the measurement results of valid events in order, represented as $a_{j}^{iz}$ $\left(a_{j}^{ix}\right)$ and $b_{j}^{iz}$ $\left(b_{j}^{ix}\right)$, $i$ is the index of $\rm Bob$ and $j$ indicates the sequence of outcomes. They classified the valid events according to the selected basis and matched the measurement results in the same basis sequentially.} Considering their common origin in the same Bell state,  $a_{j}^{iz}$ $\left(a_{j}^{ix}\right)$ and $b_{j}^{iz}$ $\left(b_{j}^{ix}\right)$ are equal without the effect of noise. 

\textcolor{black}{We provide a detailed explanation for post-matching method using an example. In a three-party scenario, measurement events between $\rm Alice$ and $\rm Bob_i (\mathit{i}\in \{1, 2\})$ can be represented as $S_{k}^{i}$, where $k$ is the sequence of measurement events. It is supposed that the first four measurement events can be denoted in sequence as $\{S_{1}^{1,X}$, $S_{2}^{1,D}$, $S_{3}^{1,Z}$, $S_{4}^{1,Z}\}$ and $\{S_{1}^{2,Z}$, $S_{2}^{2,Z}$, $S_{3}^{2,X}$, $S_{4}^{2,D}\}$. The superscript $X$ indicates that $\rm Alice$ and $\rm Bob_\mathit{i}$ both conduct measurements in the X basis, while the superscript $Z$ means that $\rm Alice$ and $\rm Bob_\mathit{i}$ both conduct measurements in the Z basis. The superscript $D$ encompasses all other cases. In this scenario, the measurement results of $S_{1}^{1,X}$ are matched with those of $S_{3}^{2,X}$, and get $a_{1}^{1x}$, $a_{1}^{2x}$, $b_{1}^{1x}$, and $b_{1}^{2x}$. Similarly, The measurement results of $S_{3}^{1,Z}$ $(S_{4}^{1,Z})$ are matched with those of $S_{1}^{2,Z}$ $(S_{2}^{2,Z})$ to get $a_{1}^{1z}$, $a_{1}^{2z}$, $b_{1}^{1z}$, and $b_{1}^{2z}$ ($a_{2}^{1z}$, $a_{2}^{2z}$, $b_{2}^{1z}$, and $b_{2}^{2z}$).}

Subsequently, $\rm Alice$ performs XOR operations on the measurement results in the Z basis: $c_{j}^{iz}= a_{j}^{1z}\oplus a_{j}^{iz}$ . After $\rm Alice$ publishes $c_{j}^{iz}$, $\rm Bob_i$ perform another XOR operation on $c_{j}^{iz}$ and $b_{j}^{iz}$ to obtain a new classical bit string ${b_{j}^{iz}}^{\prime}$. In the case of the X basis, $\rm Alice$ acquires ${a_{j}^{1x}}^{\prime}$, which is the outcomes of $a_{j}^{1x} \oplus a_{j}^{2x} \oplus \cdots \oplus a_{j}^{(n-1)x}$. Based on the self-inverse property of XOR, there are: 
\begin{equation}\label{equation1}
{b_{j}^{iz}}^{\prime}=a_{j}^{iz}\oplus a_{j}^{1z}\oplus b_{j}^{iz}.
\end{equation}
\textcolor{black}{Through the post-matching and data processing steps outlined above, $\rm Alice$ and $\rm Bob_\mathit{i}$ establish correlations within each group of outcomes. They can obtain bit strings with GHZ correlation through the information processing steps above.} There are:
\begin{equation}\label{equation2}
a_{j}^{1z}={b_{j}^{1z}}^{\prime}={b_{j}^{2z}}^{\prime}=\cdots={b_{j}^{(n-1)z}}^{\prime},
\end{equation}

\begin{equation}\label{equation3}
{a_{j}^{1x}}^{\prime}=b_{j}^{1x}\oplus b_{j}^{2x}\oplus\cdots \oplus b_{j}^{(n-1)x},
\end{equation}
without considering noise.

{\it{\bf{4.}}} $\rm Alice$ and $\rm Bob_\mathit{i}$ use the valid events in the X-basis for information leakage estimation, and the valid events in the Z basis for obtaining raw keys. The bit string for QCKA can be derived through error correction and privacy amplification~\cite{bennett1996mixed} applied to the raw keys. The specific calculation steps are outlined in Sec.~\ref{sec4}.

\textcolor{black}{In our protocol, Bell states are generated and distributed by an untrusted provider (who can be the eavesdropper), which is similar to the Ekert91~\cite{ekert1991quantum} protocol. We make no assumptions about the light source in the protocol, and the attacker can generate any state he wants. However, the measurements for all participants are what we need to assume, namely perfect Z and X bases. Since the participants randomly choose the Z and X bases for measurement, these disturbances inevitably lead to an increase in error rates, which is unavoidably detectable through error rate estimation. Therefore, our protocol is a source-independent protocol~\cite{koashi2003secure}, circumventing the source vulnerabilities present in traditional prepare-and-measure protocols. Its security can be analyzed even when the source end is controlled by an untrusted node.
}

The distribution of Bell states instead of multi-photon entanglement states contributes to a simplified system design~\cite{jennewein2000quantum,pickston2023conference}. As the number of participants increases, the only requirement is adding detection devices with a corresponding number and connecting them to the optical switch through optical fibers. This scalability allows for the flexibility to add or remove participants as needed.

\section{Security analysis}\label{sec}
\begin{figure}[t!]
  \centering
  \includegraphics[width=1.0\columnwidth]{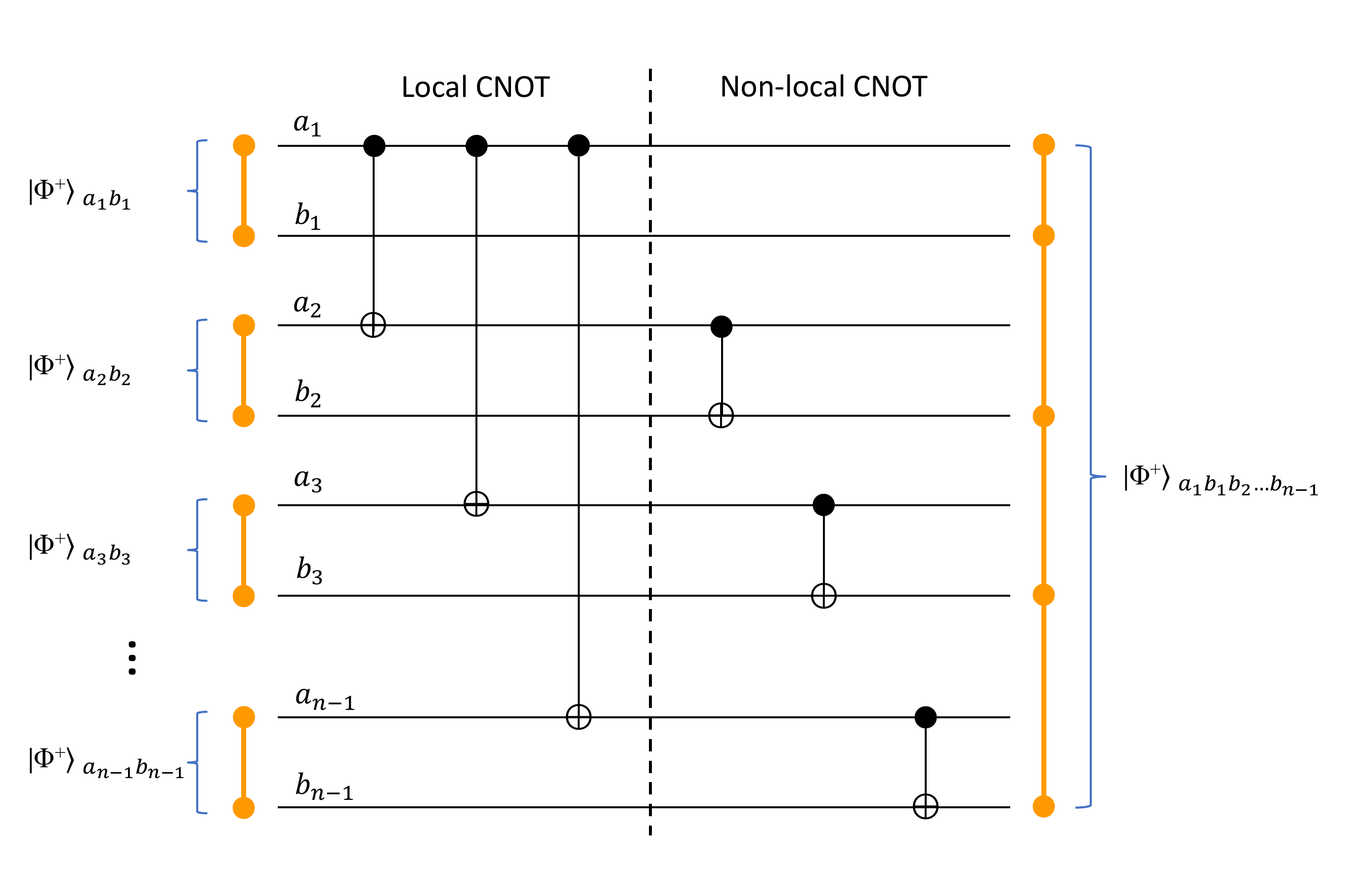}
  \caption{\textbf{GHZ state generation.}\label{fig2} $\rm Alice$ and $\rm Bob_\mathit{i}$ perform local and non-local CNOT operations on qubits they store, establishing entanglement correlations in subsystem $\left\{ a_1, b_1, b_2, \cdot\cdot\cdot, b_{n-1} \right\}$.}
\end{figure}

In this section, we introduce a virtual protocol that can be transformed into the actual QCKA protocol outlined above to demonstrate the security of our scheme. The virtual protocol allows participants to employ quantum memory for storing the generated Bell states, while the classical data storage available to participants in the actual protocol only permits them to store the results of projection measurements. The correlations among multiple Bell states stored are established by quantum gates, resulting in a different sequence of operations and measurements in the virtual protocol compared to the actual one. Figure~\ref{fig2} illustrates the generation process of GHZ states. Through multi-party entanglement purification~\cite{bennett1996mixed, lo1999unconditional, shor2000simple}, the noise introduced during the protocol can be eliminated, leading to the attainment of maximum entanglement. The specific steps of the virtual protocol are as follows:

{\it{\bf{1.}}} In the case of $n$ protocol participants, the polarization-entangled photon pair source generates $n-1$ pairs of Bell state $\left| \Phi^+ \right\rangle=\frac{1}{\sqrt{2}}(\left| 00 \right\rangle+\left| 11 \right\rangle)$ $\left(\left| \Phi^+ \right\rangle=\frac{1}{\sqrt{2}}(\left| ++ \right\rangle+\left| -- \right\rangle)\right)$. One of the qubits in Bell state is distributed to $\rm Alice$ and the other to corresponding $\rm Bob_\mathit{i}$. $\rm Alice$ and $\rm Bob_\mathit{i}$ store them by quantum memory. The total system of the generated Bell states is given by:

\textcolor{black}{\begin{equation}\label{equation5}
\left| \Phi^+ \right\rangle^{\otimes (n-1)}=\frac{1}{\sqrt{2^{n-1}}}\sum_{\substack{ a_i=b_i \\  
b_i\in \{0,1 \}}}\left| a_1b_1\cdots a_{n-1}b_{n-1}\right\rangle.
\end{equation}}

{\it{\bf{2.}}} $\rm Alice$ and $\rm Bob_\mathit{i}$ establish entanglement among the qubits they store by performing multiple CNOT operations. Taking $a_1$ as control bits, $\rm Alice$ perform local CNOT operations with $a_2, a_3\cdots a_{n-1}$ in order. Following this, the participants perform non-local CNOT operations sequentially on $a_i$ and corresponding \textcolor{black}{$b_i (i\in \{2, 3, \cdots, \mathit{n}-1\})$} as control and target bits, respectively. Without considering the noise, the evolution processes of the quantum state are as follows:

\textcolor{black}{\begin{equation}\label{equation6}
\begin{aligned}
&\frac{1}{\sqrt{2^{n-1}}}\sum_{\substack{ a_i=b_i \\  
b_i\in\{0,1 \}}}\left| a_1b_1\cdots a_{n-1}b_{n-1}\right\rangle\\
\stackrel{\rm CNOT}{\longrightarrow}&\frac{1}{\sqrt{2^{n-1}}}\sum_{\substack{ a_i=b_i \\  
b_i\in\{0,1 \}}}[\left| 00a_2b_2\cdots a_{n-1}b_{n-1}\right\rangle+ \\&
\left| 11(1 \oplus a_2)b_2\cdots (1 \oplus a_{n-1})b_{n-1}\right\rangle]\\
\stackrel{\substack{\rm non-local \\  
\rm CNOT}
}{\longrightarrow}&\frac{1}{\sqrt{2^{n-1}}}
(\left| 0 \right\rangle+\left| 1 \right\rangle)_{a_2}\cdot\cdot\cdot (\left| 0 \right\rangle+\left| 1 \right\rangle)_{a_{n-1}} \\&(\left| 00\cdot\cdot\cdot 0 \right\rangle+\left| 11\cdot\cdot\cdot 1 \right\rangle)_{a_1b_1b_2\cdot\cdot\cdot b_{n-1}}.
\end{aligned}
\end{equation}}
The state of subsystem $\left\{ a_1, b_1, b_2, \cdot\cdot\cdot, b_{n-1} \right\}$ can be expressed as:
\textcolor{black}{\begin{equation}
\left| \Phi^+ \right\rangle_{a_1b_1b_2\cdot\cdot\cdot b_{n-1}}=\frac{1}{\sqrt{2}}\left({\left| 0 \right\rangle}^{\otimes n}+{\left| 1 \right\rangle}^{\otimes n}\right)_{a_1b_1b_2\cdot\cdot\cdot b_{n-1}}.
\end{equation}}

{\it{\bf{3.}}} After obtaining a sufficient number of noisy GHZ states, the participants select a subset of them and perform measurements in the X basis to estimate parameters. Subsequently, they perform purification on the remaining GHZ states, extracting approximately pure GHZ states. The measurements of these GHZ states are performed in the Z basis. $\rm Alice$ and $\rm Bob_i$ record the results as secret keys.

The steps of virtual and actual protocols correspond one-to-one. Quantum memory corresponds to classical data storage. While quantum memory stores and retrieves quantum information using qubits, classical data storage performs a similar function in classical computing by storing and retrieving classical bits. The CNOT operations are equivalent to XOR operations and bit flips. For the Z basis, the local CNOT operations correspond to XOR operations between $a_{j}^{1z}$ and $a_{j}^{iz}$ $\left(c_{j}^{iz}= a_{j}^{1z}\oplus a_{j}^{iz}\right)$ in the actual protocol. The non-local CNOT operations correspond to deciding whether to perform a bit flip on $b_{j}^{iz}$ based on $c_{j}^{iz}$. For the X basis, the local CNOT operations correspond to that $\rm Alice$ performs computation ${a_{j}^{1x}}^{\prime} = a_{j}^{1x} \oplus a_{j}^{2x} \oplus \cdots \oplus a_{j}^{(n-1)x}$. The non-local CNOT operations do not affect the measurement outcomes in the X basis.

Entanglement purification can be converted into the quantum error
correction\cite{bennett1996mixed,shor2000simple}, where the Calderbank-Shor-Steane code are divided into two subcodes, designed for qubit flip errors and phase errors, respectively. The correction of qubit flip error and phase flip error are correspond to classical bit error correction and privacy amplification. Considering the monogamy of entanglement\cite{de2014monogamy}, these obtained pure GHZ states almost leak no information to eavesdroppers. Therefore, the actual protocol is equivalent to the virtual protocol and can effectively prevent information leakage.

\section{Results and discussion}\label{sec4}

\subsection{Calculation of the key rate}

According to the description of the protocol above, the bit strings with entanglement correlation are finally shared by the protocol participants. The key rate in the asymptotic case~\cite{fu2015long,li2023breaking} is given by: 

\begin{equation}\label{equation7}
R_{\rm QCKA}=Q_{Z}\left\{1-H\left(E_{X(n)} \right)-f {\max_{i}}\left[H\left(E_Z^{1,i}  \right)\right]\right\},
\end{equation}
where $Q_{Z}$ is the gain of the system in the Z basis, $E_{X(n)}$ is the total bit error rate in the X basis, and $E_Z^{1,i}$ indicates the marginal error rates between $\rm Alice$ and corresponding $\rm Bob_\mathit{i}$. $f$ represents the error correction efficiency, and $H(x)=-x\log_2x-(1-x)\log_2(1-x)$ is binary Shannon entropy function.

We denote the efficiency of all detectors as $\eta$. Considering the central symmetric network architecture, the quantum channels efficiency of $\rm Alice$ and $\rm Bob_\mathit{i}$ can be gotten: $\eta_A=\eta_B=\eta \times {10}^\frac{-\alpha L}{10}$, where $\alpha$ is the optical fiber attenuation coefficient, and $L$ is the length of optical fiber between participants and the untrusted entanglement provider. Taking into account the type-II source of polarization entanglement, the gain for $k$-photon pairs~\cite{kok2000postselected,ma2007quantum} is as follows:
\begin{equation}
\begin{split}
Q_k &= \left[1-(1-Y_{0A})(1-\eta_A)^k\right] \\
&\left[1-(1-Y_{0B})(1-\eta_B)^k\right] \frac{(k+1)\lambda^k}{\left(1+\lambda\right)^{k+2}}.
\end{split}
\end{equation}
By summing up $Q_k$, we can get the gain in the Z basis of two-party entanglement system: 
\begin{equation}\label{equation8}
\begin{split}
Q_Z = \sum_{k=0}^{\infty}Q_k = 1 &- \frac{1-Y_{0A}}{\left(1+\eta_A\lambda\right)^2} 
- \frac{1-Y_{0B}}{\left(1+\eta_B\lambda\right)^2} \\
&+ \frac{\left(1-Y_{0A}\right)\left(1-Y_{0B}\right)}{\left(1+\eta_A\lambda+\eta_B\lambda-\eta_A\eta_B\lambda\right)^2},
\end{split}
\end{equation}
where $\lambda$ is half of the expected photon pair number $\mu$, and $Y_{0A}$ $(Y_{0B})$ is the detector dark count. Due to the entangled photon pair source, we only need to consider the collection efficiency of two participants involved. The gain of the system \textcolor{black}{remains} unchanged with variations in the number of participants. Considering the symmetry between measurements in the Z and X bases, we can derive the bit error rate $e$ in them: 
\begin{equation}\label{equation9}
\begin{split}
e=&e_0-\\
&
\frac{2\left(e_0-e_d\right)\eta_A\eta_B\lambda\left(1+\lambda\right)}{Q_{Z}\left(1+\eta_A\lambda\right)\left(1+\eta_B\lambda\right)\left(1+\eta_A\lambda+\eta_B\lambda-\eta_A\eta_B\lambda\right)},
\end{split}
\end{equation}
where $e_0$ is the background error rate, $e_d$ is the misalignment rate. $e$ represents the probability that $\rm Alice$ and $\rm Bob_\mathit{i}$ have discordant outcomes in a single valid event.

\begin{figure}
    \centering
     
    \begin{subfigure}{0.5\textwidth}
        \centering
        \includegraphics[width=1\textwidth]{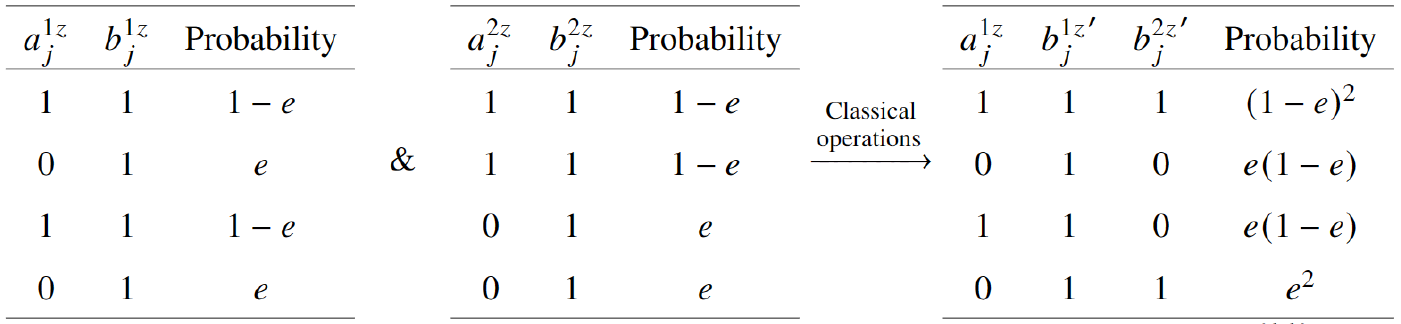}
        \caption{}
        \label{fig:a}
    \end{subfigure}

    \begin{subfigure}{0.5\textwidth}
        \centering
        \includegraphics[width=1\textwidth]{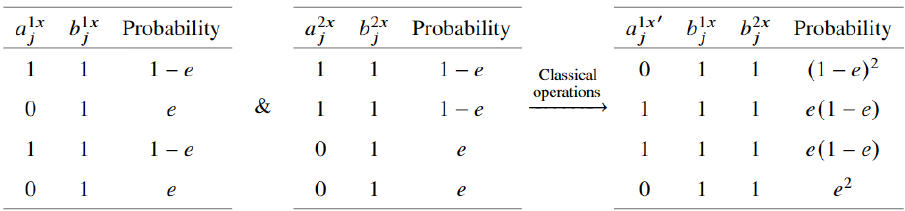}
        \caption{}
        \label{fig:b}
    \end{subfigure}
    \caption{\textbf{The error rate analysis.} The probabilities of various outcomes arising after measurement and classical operations are provided. (a) The second and fourth row of the table represent the marginal error rate $E_Z^{1,1}$. The third and fourth row represent the marginal error rate $E_Z^{1,2}$. (b) The second and third row represent the total error rate $E_{X(3)}$, which is the probability that measurement results fail to satisfy entanglement correlation.}
    \label{figt}
\end{figure}

To estimate $E_{X(n)}$ and $E_Z^{1,i}$, we present tables in Fig.~\ref{figt} for the mesurement results in three-party scenario and extended the analysis to multiple participants, assuming that no errors are introduced during the classical operations. According to Fig.~\ref{fig:a} the generation of a correct key becomes unattainable if any one measurement result in the Z basis contains an error. Since the marginal error rates $E_Z^{1,i}$ are exclusively introduced by the discordant outcomes of $\rm Alice$ and $\rm Bob_\mathit{i}$ in a single valid event, there are $E_Z^{1,i} = e(1 - e) + e^2 = e$. which remains unaffected by variations in the number of participants. The total bit error rate of $n$ participants where at least one measurement result of $\rm Bob_\mathit{i}$ differs from $\rm Alice's$ in the Z basis is as follows:

\begin{equation}\label{equation10}
E_{Z(n)}=1-\left(1-e\right)^{(n-1)}.
\end{equation}

While entangled photon pairs are measured in the X basis, the data in the first and fourth rows in Fig.~\ref{fig:b} satisfy Eq.~\ref{equation3}. Therefore, in the three-party scenario, the total bit error rate in the X basis is denoted as $E_{X(3)}=2e(1-e)$. Considering the identity property of the XOR operation, with an even number of incorrect measurements, the correctness of the results will not be affected. In this case, the key distribution is considered to be successful. Otherwise, it is deemed a failure. In the expansion of $E_{X(n)}$, there are no even-powered terms of $e$, while odd-powered terms remain unaffected:

\begin{equation}\label{equation11}
E_{X(n)}=\sum_{i=0}^{t}\mathrm{C}_{n-1}^{2i+1}\left(1-e\right)^{n-2i-2}e^{2i+1}.
\end{equation}
When $n$ is an even number, $t=\frac{n}{2}-1$. When $n$ is an odd number, $t=\frac{n-3}{2}$.

The finite key analysis for the experiment is provided below. The universally composable security framework is employed as our security criteria. We define the length of the security key as $L_{\rm QCKA}$ according to the calculation in Ref.~\cite{yin2020tight,grasselli2018finite,li2023breaking}: 
\begin{equation}\label{equation12}
\begin{split}
L_{\rm QCKA} &= n_z \left\{1-H({\phi^z})-f \max_i \left[H\left(E_Z^{1,i}  \right)\right]\right\} \\
&\quad - \log_2{\frac{2(n-1)}{\varepsilon_{\rm cor}}} - 2\log_2{\frac{1}{2\varepsilon_{\rm sec}}},
\end{split}
\end{equation}
where $n_z$ is the number of entangled pairs measured in the Z basis. \textcolor{black}{$\phi^z$ is the upper limit of the phase error rate in the Z basis considering statistical fluctuations. }$\varepsilon_{\rm cor}$ is the failure probability of error verification, and $\varepsilon_{\rm sec}$ is the failure probability of privacy amplification~\cite{bai2022post}.

\begin{figure}[t!]
  \centering
\includegraphics[width=1\columnwidth]{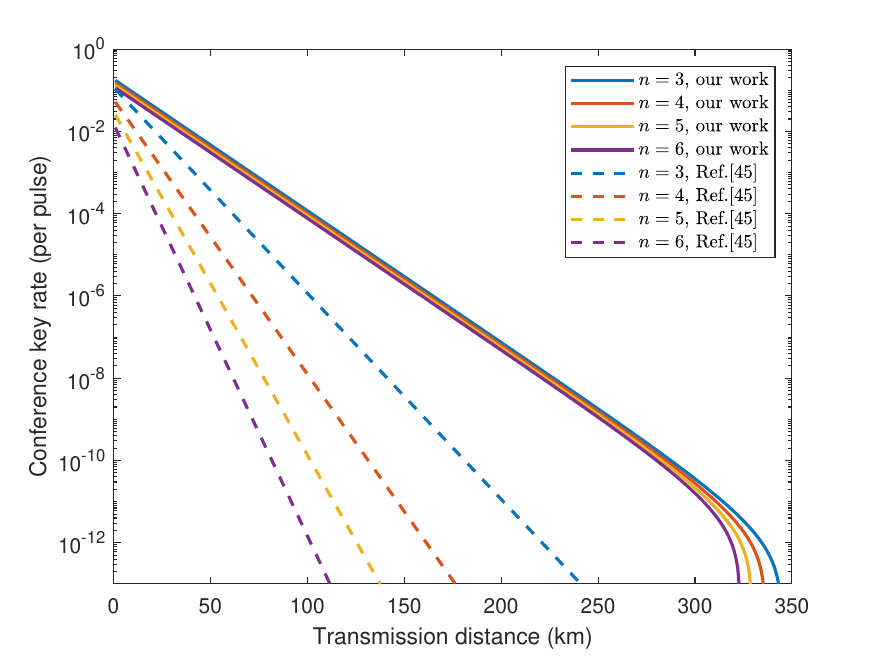}
  \caption{\textbf{Key rate of our protocol and N-BB84.}\label{fig4} Under the conditions of perfect entanglement sources, we numerically simulate the key rate for different number of participants (3, 4, 5, 6) using the parameters in Table~\ref{table1}. The horizontal axis represents the distance between the central node and each participants. The solid line represents key rate of our protocol, while the dashed line represents that of N-BB84~\cite{grasselli2018finite} protocol.
  }
\end{figure}

\begin{table}[h]

\centering

\caption{\textbf{Simulation parameters.}\label{table1} $e_0$ is the background error rate. $e_d$ is the misalignment rate. $\eta_d$ is the detection efficiency of single photon detectors. $p_d$ is the dark count rate. $\alpha$ is the attenuation coefficient of the ultra-low-loss fiber. $f$ is the error correction efficiency. $\varepsilon_{\rm cor}$, $\varepsilon_{\rm sec}$ are the parameters of correctness and privacy.}

\begin{tabular}{cccccccc}
\\
\hline\hline
$e_0$ & $e_d$ & $\eta_d$ & $p_d$ & $\alpha$ & $f$&$\varepsilon_{\rm cor}$& $\varepsilon_{\rm sec}$ \\
\hline
 $0.5$ & $0.02$ & $56\%$ & $10^{-7}$  & $0.16$ & $1.16$ & $1.2 \times 
 10^{-9}$ & $1.2 \times 
 10^{-9}$\\
\hline\hline
\end{tabular}
\end{table}

\begin{figure}[t!]
  \centering
\includegraphics[width=1\columnwidth]{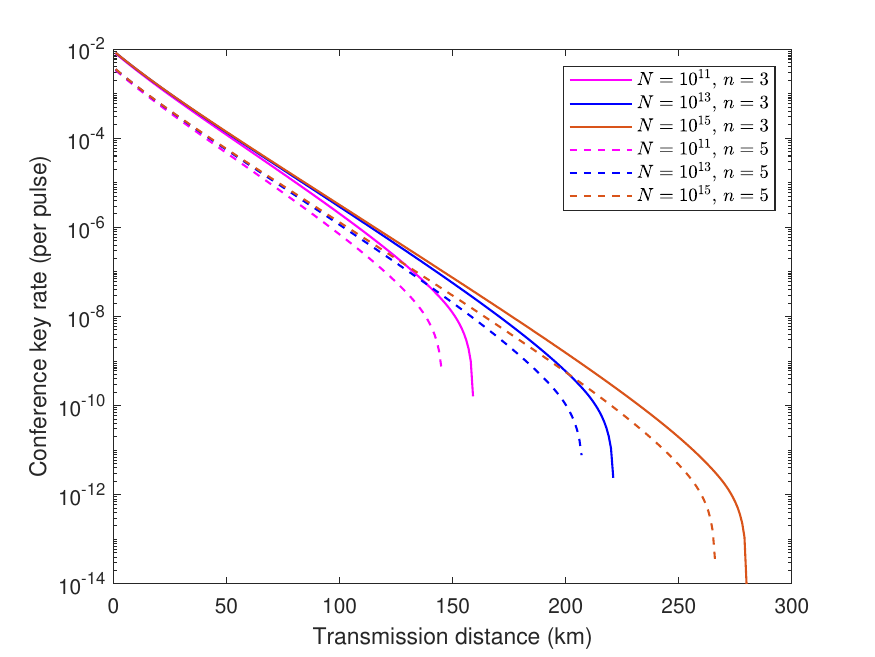}
  \caption{\textbf{Simulation schematic under finite key conditions.}\label{fig5} 
The solid lines represent the key rates corresponding to different key lengths in the case of three participants, while the dashed lines indicate the key rates in the scenario of five participants. $N$ is the number of pulses allocated to each participant by the entanglement source.}
\end{figure}

\textcolor{black}{
\begin{figure}[t!]
  \centering
\includegraphics[width=1\columnwidth]{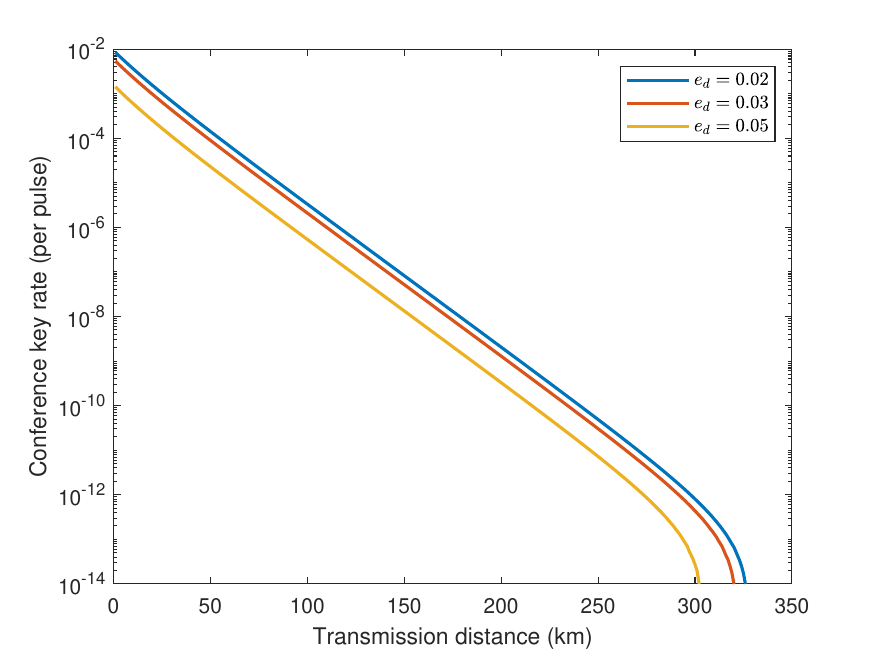}
\caption{\textbf{Simulation schematic under different misalignment rate of the basis.}\label{fig6} We provide the asymptotic key rate under the scenario of three participants. Our protocol can still achieve a transmission distance exceeding 300 kilometers with a misalignment rate of 0.05 for the basis.}
\end{figure}}

\textcolor{black}{Due to the challenge of counting phase errors in valid events, it is not feasible to obtain the phase error rate in the Z basis directly. However, since the density matrix of the n-photon state is the same in the Z and X bases, bit flips in the X basis correspond to phase flips in the Z basis. Consequently, we can utilize the property that the phase error rate in the Z basis is identical to the bit error rate in the X basis to calculate $\phi^z$.}

\textcolor{black}{We define the superscript $*$ and overline to denote expected value and observed value.} The number of bit error under the X basis can be defined as $m_x=NQ_XE_{X(n)}$, assuming that the gain in the \textcolor{black}{X basis} $Q_X=Q_Z$. Then the numerical equation for the variant of Chernoff-bound~\cite{chernoff1952measure,yin2020tight} is used to obtain the limit of expected value $m_x^*$. $m_x^*$ = $m_x+\beta+\sqrt{2\beta{m_x}+{\beta}^2}$, of which $\beta=-\ln\varepsilon$ and $\varepsilon = 10^{-10}$ is the failure probability. \textcolor{black}{Since the phase error rate in the Z basis is identical to the bit error rate in the X basis in the asymptotic limit, there is $m_{zt}^*=m_x^*n_z/n_x$,} where $m_{zt}^*$ represents \textcolor{black}{the upper limit of the expected number} of qubits with phase errors in the Z basis. 
Subsequently, a similar procedure is employed to estimate the upper limit of the number of phase errors observed in the Z basis: ${\bar{m}_{zt}}={m_{zt}^*}+\frac{\beta}{2}+\sqrt{2\beta{m_{zt}^*}+\frac{\beta^2}{4}}$. Ultimately, the upper limit of the phase error rate can be determined as \textcolor{black}{$\phi^z={\bar{m}_{zt}}/n_z$}. 

\subsection{Simulation results} 

The convex optimization algorithm is used to simulate the protocol. We provide the asymptotic key rates for both our protocol and N-BB84 in Fig.~\ref{fig4}. For a fair comparison, it is assumed that perfect entanglement sources are used in the system, which means a guarantee of perfectly entangled photon pairs. The specific parameters used are shown in Table~\ref{table1}. The results shows that under identical experimental parameters and post-processing procedures, our protocol consistently exhibits a higher key rate and achieves a longer transmission distance. At a transmission distance of 200 km, the key rate of our protocol can be three orders of magnitude higher than that of N-BB84. The advantage becomes more evident with an increasing participant number. Our protocol still achieves successful key generation with 6 participants at 320 km. The robustness of our protocol in the face of rise in the number of participants is attributed to the improvement of valid event probability. Therefore, the protocol's key rate is enhanced from$O(\eta^{n})$ to $O(\eta^{2})$, making it more suitable for large-scale quantum networks compared to previous entanglement-based QCKA protocols.

We have also calculated key rate under different constraint conditions taking the imperfect source into consideration. Figure~\ref{fig5} shows the key rate at different distances under finite-key conditions when the number of participants is 3 and 5, which provides a reference for the application of protocol. Under the condition of five participants, our protocol maintains the ability to generate keys at a distance of 140 km when $N=10^{11}$. Figure~\ref{fig6} illustrates that in the case of three participants, the key rate under different values of $e_d$. The transmission distance can reach more than 300 km with a basis misalignment rate of $5\%$, which means it can withstand more interference from the field environment.

\section{Conclusion}\label{con}

We have proposed a QCKA scheme based on Bell states, which can be extended to an arbitrary number of participants. The protocol establishes correlations among the measurement results of Bell states through classical operations and the post-matching method, thereby avoiding the challenges in generating multi-photon entangled states. A performance comparison with N-BB84 and a finite key analysis, considering composable security, is provided. Simulation results show that our protocol achieves a transmission distance exceeding 320 kilometers in the case of 6 participants and can tolerate a $5\%$ misalignment rate of the basis, which represents an improvement over existing entanglement-based protocols. \textcolor{black}{In recent years, significant progress has been made in the advantage distillation technique~\cite{li2022improving,zhang2024discrete} within the realm of quantum key distribution. This technique enhances the correlation between raw keys, allowing for greater tolerance to channel loss and error rates. It's foreseeable that advantage distillation will be widely applied in QCKA. We believe that incorporating this technique into our protocol can further enhance its performance.}

The flexibility and scalability of our protocol enable the free selection of new participants from network users, without requiring hardware changes for existing participants and entanglement providers. Therefore, our protocol can be inserted into the entanglement network with plug-and-play capability. Additionally, it constitutes a crucial application of the post-matching method, indicating greater adaptability to multi-party applications. Similar to Ref.~\cite{liu2023experimental}, our protocol requires only minor adjustments to basis selection probabilities and post-processing procedures for efficient execution of quantum secret sharing tasks. This suggests the potential for extending our protocol to other areas of quantum cryptography.

\section*{Acknowledgments}
This work was supported by the National Natural Science Foundation of China (No. 12274223); Program for Innovative Talents and Entrepreneurs in Jiangsu (No. JSSCRC2021484) and Program of Song Shan Laboratory(included in the management of Major Science and Technology Program of Henan Province) (No. 221100210800-02).


%

\end{document}